\documentclass{article}

\usepackage[round]{natbib}
\usepackage{amsmath,amssymb,amsthm,bm,enumerate,mathrsfs,mathtools}
\usepackage{latexsym,color,verbatim,multirow}
\usepackage{graphicx}
\usepackage{caption}
\usepackage{subcaption}
\usepackage{enumitem}

\newcommand{\Nu}{{\cal V}}
\def\be {{\bf 1}}
\newcommand{\intercept}{\alpha}

\newtheorem{theorem}{Theorem}

\newtheorem{lemma}[theorem]{Result}

\begin{document}

\title{Post-selection inference\\ for $\ell_1$-penalized likelihood models}
\author{Jonathan Taylor and  Robert Tibshirani\\
Stanford University } 

\maketitle

\begin{abstract}
We present a new method for post-selection inference for  $\ell_1$ (lasso)-penalized likelihood models,
including generalized regression models.   Our  approach generalizes the post-selection
framework presented in 
\citet{lee2014}. 
The method  provides p-values and  confidence intervals that are asymptotically
valid, conditional on the inherent selection done by the lasso. We present applications of this work to   (regularized)
logistic regression, Cox's proportional hazards model and the graphical lasso.
We do not provide rigorous proofs here of the claimed results, but rather conceptual  and  theoretical sketches.

\end{abstract}

\section{Introduction}
\label{sec:intro}
Significant recent progress has been made in the problem of inference after selection for Gaussian regression models.
In particular,
\citet{lee2014} 
derives closed form p-values and confidence intervals, after fitting the lasso with a fixed value
of the regularization parameter, and 
\citet{spacing}
provides analogous results for forward stepwise regression and least angle
regression (LAR). In this paper we derive a simple and natural way to extend these results to $\ell_1$-penalized likelihood models,
including generalized regression models  such as (regularized)
logistic regression  and Cox's proportional hazards model. 

Formally, the results described here are contained  in 
\cite{TT2015}
in which the authors consider the same problems but having  adding noise to the data  before fitting the model and carrying out the selection. Besides
expanding on the GLM case, considered only briefly in
\cite{TT2015},
the novelty in this work is the second-stage
estimator, which is asymptotically equivalent to the post-LASSO MLE, overcomes some problems encountered with a second-order remainder.
And unlike the proposals in 
\cite{TT2015},
 the estimator proposed here does not require MCMC sampling and is computable in closed form. 
Our estimator comprises of a single step of Newton-Raphson (or equivalently Fisher scoring) in the {\em selected model} after having fit the LASSO.
This is discussed further in Section \ref{sec:theory}. 
We note that one-step estimators are commonly used in semi-parametric inference: see for example \cite{bickel1993}.

In this paper we do not provide rigorous proofs of the claimed results, but rather theoretical and conceptual sketches, together with numerical evidence.
We are confident that rigorous proofs can be given (with appropriate assumptions) and plan to report these elsewhere.
We also note the strong similarity between  our one-step estimator  and the ``debiased lasso'' construction of \cite{zhangconf}, \cite{buhlsignif},  \citet{vdgsignif}, and \citet{montahypo2}. This connection is detailed in Remark A of this  paper.

\begin{figure}[ht]
\includegraphics[width=\textwidth]{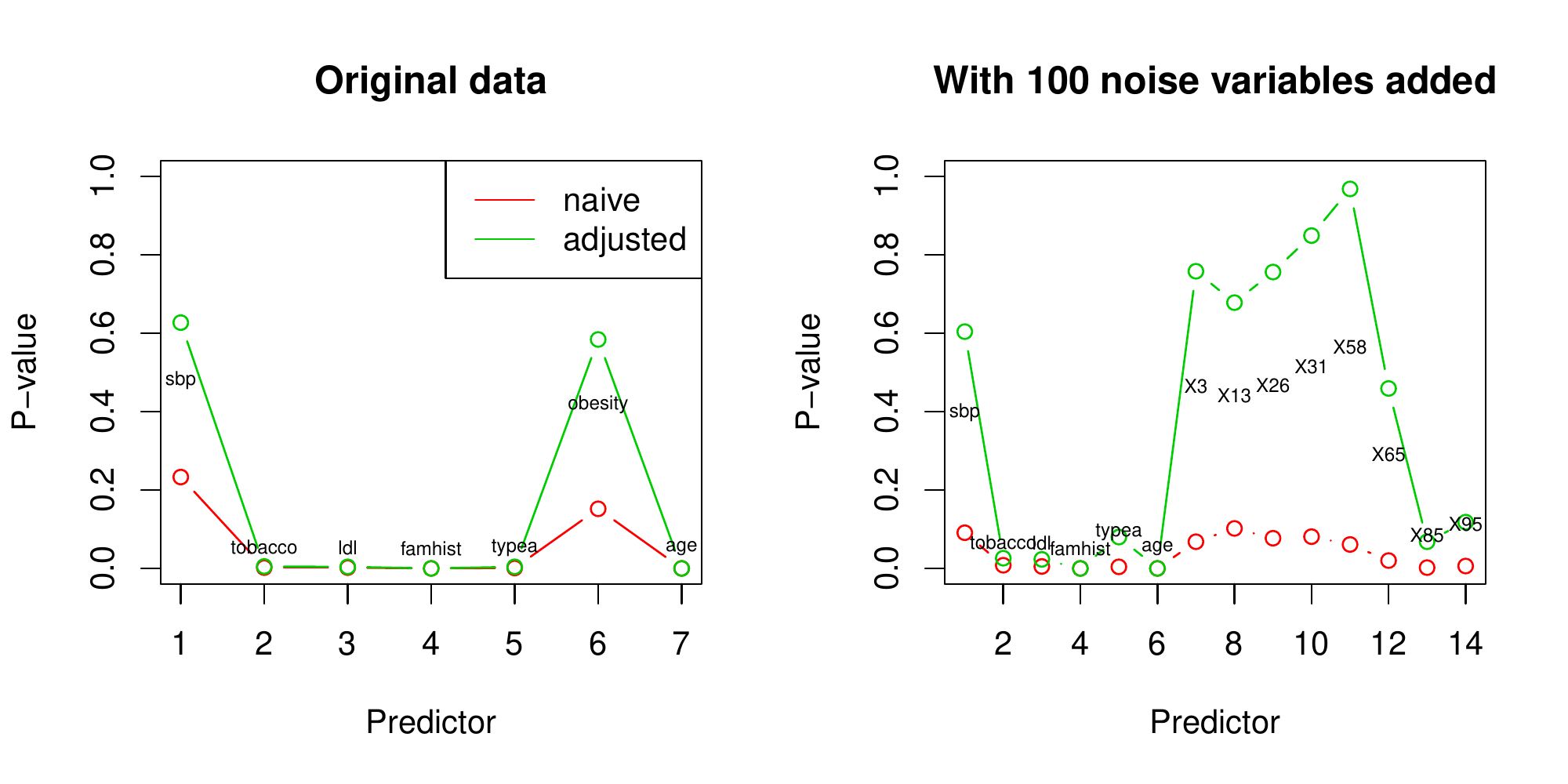}
\caption{\em South African Heart disease data. P-values from naive and selection-adjusted approaches, for original data (left) and  data with
100 additional noise predictors (right). Each model was chosen by lasso-penalized logistic regression, choosing the tuning parameter by cross-validation.}
\label{fig:fig1}
\end{figure}

Figure \ref{fig:fig1} shows an example--- the South African heart disease data.
These are a retrospective sample of 463 males in a heart-disease high-risk region
of the Western Cape, South Africa. The outcome is binary--- coronary heart disease--- and there are 9 predictors.
We applied lasso-penalized logistic regression, choosing the tuning parameter by cross-validation.
The left panel shows the standard  (naive)  p-values and the post-selection p-values from our theory, for the predictors in the active set.
Since the sample size is large compared to the number of predictors, the unadjusted and adjusted p-values are only substantially different
for two of the predictors.
On the right we have added 100 independent standard Gaussian predictors  (labelled $X1, X2 \ldots  X100$) to examine the effects of selection. 
Now the naive p-values are unrealistically small for the noise variables while the adjusted p-values are appropriately large.
Although our focus  is on selection via $\ell_1$-penalization, a similar approach can likely be applied to forward stepwise
methods for likelihood models, and in principle, least angle regression (LAR) though algorithms for LAR in the generalized
linear model setting are less developed with \cite{glmpath} one exception. 

An outline of this paper is as follows. Section \ref{sec:postsel}  reviews post-selection inference for the lasso in the Gaussian regression model
and introduces 
our proposal for more general (non-Gaussian) generalized linear models. In Section \ref{sec:theory} we give an equivalent form of the proposal, one that
applies to general likelihood models, for example the graphical lasso. We give a rough argument for the asymptotic validity of the procedure.
Section \ref{sec:simulations} reports a simulation study of the methods. In Section \ref{sec:examples}
we show an example of  the proposal applied to Cox's model for survival data. 
The graphical lasso is studied in Section \ref{sec:glasso}.
We end with a discussion in Section \ref{sec:discussion}.

\section{Post-selection inference for generalized regression models}
\label{sec:postsel}
Suppose  that we have data $(x_i, y_i), i=1,2,\ldots N$ consisting of features $x_i=(x_{i1},x_{i2},\ldots x_{ip})$ and outcomes $y_i, i=1,2,\ldots N$.
Let $X= \{x_{ij}\}$ be the $N\times p$  data matrix. 
We consider a generalized regression model  with linear predictor $\eta=\intercept+x^T\beta$ and log-likelihood
$\ell(\intercept,\beta)$.  Our   objective function
has the form\begin{eqnarray}
J(\intercept,\beta)=-\ell(\intercept,\beta) +\lambda\cdot \sum_1^p |\beta_j| 
\label{eqn:obj}
\end{eqnarray}

 Let $\hat\intercept, \hat\beta$ be the minimizers of $J(\intercept, \beta)$. We wish to carry out selective inference
for some functional  $\gamma^T\beta$. For example, $\gamma$  
might be chosen so that   $\gamma^T\beta$ is the population  partial regression coefficient for the $j$th predictor. 

As a leading example, we consider  the  logistic regression model  specified by
\begin{equation}
\label{eq:logistic:model}
\pi=E(Y|x);\;  \log \pi/(1-\pi)=\intercept+x^T\beta;
\end{equation}
$$ \ell(\intercept,\beta)=\sum [y_i \log(\pi_i)+(1-y_i)\log(1-\pi_i)]$$
Having fit this model using a fixed value of $\lambda$, we carry out  post-selection inference, as in the heart disease example above. 

The reader may well
ask: ``partial regression coefficient with respect to what'?', i.e. what other covariates are we going to control for? In this paper, we follow
the selected model framework described in 
\cite{fithian14:_optim_infer_after_model_selec}
 so that having observed 
${M}$, the sparsity pattern of $\hat{\beta}$ as returned by the LASSO, we carry out selective inference for linear functionals of $\beta_M \in \mathbb{R}^M$
under the assumption that the model \eqref{eq:logistic:model} is correct with parameter $\beta^*$ satisfying $\beta^*_{-M}=0$. That is,
we carry out selective inference under the assumption that the LASSO has screened successfully, at least approximately.

There are various ways we might modify this model, though
we only consider mainly the parametric case here. For instance,
we might assume that $y$ is conditionally independent of $X_{-M}$ given $X_M$,  but  not assume the correctness of the
logistic link function.  In this case, 
the covariance matrix of our limiting Gaussian distribution (described below) is not correct, and the asymptotic theory in Section \ref{sec:theory} should be modified by using a consistent estimator
of the covariance matrix.
Alternatively, we might wish to be robust to the possibility that $X_{-M}$ may have some effect on our sampling distribution, which would also change the limiting
Gaussian distribution that we use for inference. For this reason,
the results in Section 4.2 of 
\cite{TT2015}
use bootstrap or jackknife to estimate the appropriate covariance in the limiting Gaussian distribution.
A short discussion of how this can be done is given in Section \ref{sec:randomX}.
While robustness to various mis-specifications are important issues, in this paper we focus mainly on the simpler case of providing
inference for parameters of \eqref{eq:logistic:model} under the assumption that the model chosen by the LASSO screens, i.e. has found a superset of the
true variables.

\subsection{Review of the Gaussian case}
For background, we first review the Gaussian case $y\sim N(\mu, I\cdot\sigma^2)$, developed in 
\citet{lee2014}.
We denote the selected model by $M$ with sign vector $s$.
Assuming the columns of $X$ are in general position, the  KKT conditions \cite{RyanTibshirani2012} state that $\{\hat M,\hat s \} = \{M,s_M\}$ if and only if there exists $\hat{\beta}_M \in \mathbb{R}^M$ and $u_{-M} \in \mathbb{R}^{-M}$ satisfying   
\begin{eqnarray}
X_M^T(X_M\hat{\beta}_M- y) +\lambda s_M&=&0 \cr
X_{-M}^T(X_M \beta_M-y)+ \lambda u_{-M}&=&0 \cr
{\rm sign}(\hat{\beta}_M)&=&s_M \cr
||u_{-M}||_\infty &\leq & 1 
\end{eqnarray}
This allows us to write the set of responses $y$ that yield the same $M$ and $s$  in the polyhedral form 
\begin{equation}
\{Ay\leq b\}
  \end{equation}
  where the matrix $A$ and vector $b$ depend on $X$ and the selected model, but not on $y$.
  Let $P_M$ is the orthogonal
projector onto the model subspace.
Due to the special form of the LASSO optimization problem, it turns out that the rows of $A$ (and $b$) can be partitioned so that
we can rewrite the above as
\begin{equation}
\label{eq:lasso:poly}
\{A_1\hat{\beta}_M(y)\leq b_1, A_2(I - P_M)y \leq b_2\}
  \end{equation}
where $\hat{\beta}_M(y) = (X_M^TX_M)^{-1}X_M^Ty$ are the usual OLS estimators of $\beta_M$ and $(I-P_M)y$ are the usual OLS residuals. 

This result can be used to make conditional inferences about any linear functional $\gamma^T\mu$, which we assume satisfies $P_M\gamma=\gamma$.
This assumption is roughly equivalent to assuming that we are interested in a linear functional of $P_M\mu$.
By conditioning on $P_{\gamma^\perp }y$ we obtain the
exact result based on truncated Gaussian distribution
\begin{eqnarray}
\label{eq:tgauss}
F_{\gamma^T\mu,\; \sigma^2\gamma\|_2^2}^{\Nu^-,\Nu^+}\left(\gamma^T y\right)\;|\;\{A y\leq b\}\sim {\rm U}(0,1).
\label{eq:tn0}
\end{eqnarray}
Expressions for $A, b$ and the truncation limits  $\Nu^-, \Nu^+$ are given in 
\citet{lee2014}
and  are reproduced here in the Appendix.
The relation \eqref{eq:lasso:poly} implies that the result \eqref{eq:tgauss} holds even if we condition only on $(P_M - P_{\gamma})y$, i.e. the variation
of $y$ within the model. This follows since the second condition in \eqref{eq:lasso:poly} is independent of the first condition,  and is fixed
after conditioning on $P_{\gamma^\perp}(y)$
The difference between these two conditional distributions really depends on which model we are interested in. 
We refer to conditioning on $P_{\gamma}^{\perp}y$ as inference in the saturated model, i.e. the collection of distributions
$$
\left\{N(\mu, \sigma^2 I): \mu \in \mathbb{R}^n, \sigma^2 > 0\right\}.
$$ 
We refer to conditioning on $(P_M - P_{\gamma})y$ as inference in the selected model. Formally speaking, 
we define the selected model as follows: given a subset of 
variables $E$, the selected model corresponding to $E$ is the collection of distributions
\begin{equation}
\label{eq:selected:model}
\left\{N(X_E\beta_E, \sigma^2 I): \beta_E \in \mathbb{R}^E, \sigma^2 > 0 \right\}.
\end{equation}
This distinction is elaborated on in 
\citet{lee2014}.
In this work, we only consider inference under \eqref{eq:selected:model} where the subset of variables $E$ are those
chosen by the LASSO. 
In principle, however, a researcher can add or delete variables from this set at will if they make their
decisions based only on the set of variables chosen by the LASSO. This changes the distributions for inference, meaning that
the analog of \eqref{eq:tgauss}  may no longer be the correct tool for inference.

\subsection{Extension to generalized regression models}

In this section, we  make a parallel between the Gaussian case and the generalized linear model setting. 
This parallel should be useful to statisticians familiar with the usual iteratively reweighted least squares (IRLS)
algorithm to fit the unpenalized logistic regression model. While this parallel may be useful, the formal justification is
given in Section \ref{sec:theory}. The method used to solve the optimization
problem is unrelated to the results presented in Section \ref{sec:theory}.

A common strategy for minimizing (\ref{eqn:obj}) is to express the usual
Newton-Raphson update as an IRLS step, and then
replace the weighted least squares step by a penalized  weighted least squares
procedure. For simplicity, we assume $\intercept=0$ below, though our formal justification
in Section \ref{sec:theory} will allow for an intercept as well as other
unpenalized features.

In detail,  recalling that ${\ell}$ is the log-likelihood, we define
$$W=W(\beta)=  -\Bigl(\frac{\partial^2 {\ell}} {\partial\eta\eta^T}\Bigr)\biggl|_{\eta=X\beta}$$ 
and $$z=z(\beta)=X\beta+W^{-1} \Bigl(\frac{\partial\ell}{\partial\eta }\Bigr)\biggl|_{\eta=X\beta}.$$
Of course, in the Gaussian case, $W=I$ and $z=y$.

In this notation, the Newton-Raphson step (in the unpenalized regression model) from a current value $\beta_c$ can be expressed as 
\begin{eqnarray}
\text{minimize}_{\beta} \frac{1}{2} (z(\beta_c)- X\beta)^T W(\beta_c)(z(\beta_c)-X\beta).
\label{taylor}
\end{eqnarray}
In the $\ell_1$ penalized version, each minimization has an $\ell_1$ penalty attached.

To  minimize (\ref{eqn:obj}), IRLS proceeds as follows:
\begin{enumerate}[itemsep=0mm]
\item Initialize $\hat\beta=0$
\item Compute $W(\hat{\beta})$ and $z(\hat{\beta})$ based on the current value of $\hat\beta$
\item Solve 
$$\text{minimize}_{\beta} \frac{1}{2}(z-X\beta)^TW (z-X\beta)+\lambda\cdot\sum|\beta_j|.$$ 
\item Repeat steps (2) and (3) until $\hat\beta$ doesn't change more than some pre-specified threshold.
\end{enumerate}

In  logistic regression for example, the specific forms of the relevant quantities are 
 \begin{eqnarray}             
z&=&\hat\intercept+X\hat\beta+(y-\hat \pi) \cdot {\rm Diag}[  1/(\hat \pi(1-\hat \pi))]\cr
W&=&{\rm diag}( \hat \pi_i(1-\hat \pi_i))
\end{eqnarray}
Another example is Cox's proportional hazards for censored survival data. The partial likelihood estimates
can again be found via an  IRLS procedure. More generally, both of these
examples are special cases of the one-step estimator 
described in Section \ref{sec:theory} below. The details of the adjusted dependent variable $z$ and weights $w$ can be found, for example, in
\citet{HT90} Chapter 8, pp. 213--214). We give examples of both of these applications later.

How do we carry out post-selection inference in this setting?
Our  proposal is to treat the final iterate as a weighted least squares regression, and  hence  use the approximation 
\begin{equation}
z\sim N(\mu,W^{-1}).
\label{eq:zform}
\end{equation}
 Using this idea, we simply apply the polyhedral lemma to the region $\{A z \leq b\}$ (see the Appendix).
A potential problem with this proposal is that $A$ and $b$ depend on $\hat\beta$ and hence on $y$. As a result, the  region $A z \leq b$  does not exactly correspond to the values  of  the response vector $y$ yielding the same active set and signs as our original fit.
The other obvious problem is that
of course $z$ is not actually normally distributed. 
Despite these points, we  provide evidence in Section \ref{sec:theory}  that this procedure yields asymptotically correct inferences.

\subsection{Details of the procedure}

Suppose that we have iterated the above procedure until we are at a fixed point.
The ``active'' block of the stationarity conditions has the form 
\begin{equation}
X_M^T W(z-X_M\hat{\beta}_M) = \lambda s_{M}
\end{equation}
where $W=W(\hat{\beta}_M), z=z(\hat{\beta}_M)$. 
Solving for $\hat{\beta}_M$ yields
$$\hat\beta_M=(X_M^TW X_M)^{-1}(X_M^TW z-\lambda s_M).$$
Thinking of $z$ as analogous to $y$ in the Gaussian case, this equality can be re-expressed as
\begin{equation}
\label{eq:onestep0}
\bar{\beta}_M \equiv (X_M^TW X_M)^{-1}X_M^TW z = \hat{\beta}_M + \lambda (X_M^TWX_M)^{-1} s_M.
\end{equation}
Note that  $\bar{\beta}_M$ solves
$$
X_M^T W(z-X_M\bar{\beta}_M) = 0.
$$
This last equation is {\em almost} the stationarity conditions of the unpenalized MLE for the logistic
regression using only the features in $M$. 
The only difference is that above,  the  $W$ and $z$ are evaluated at
$\hat{\beta}_M$ instead of $\bar{\beta}_M$.

Ignoring this discrepancy for the moment, recall that the active block of the
KKT conditions in the Gaussian case can be expressed in terms of the usual OLS
estimators \eqref{eq:lasso:poly}. 
This suggests the correct analog of the ``active'' constraints:
\begin{equation}
\label{eq:logistic:constraint}
\left\{y: \text{sign}(\bar{\beta}_M(y) - \lambda (X_M^TWX_M)^{-1} s_M) =s_M \right\}.
\end{equation}

Let's take a closer look at $\bar{\beta}_M$:
$$
\begin{aligned}
\bar{\beta}_M &= (X_M^TW X_M)^{-1}(X_M^TW z) \\ 
&= (X_M^TW X_M)^{-1}\left(X_M^TW X_M\hat{\beta}_M + \frac{\partial}{\partial \beta_M} \ell_M(\beta_M) \biggl|_{\beta_M=\hat{\beta}_M} \right) \\
&= \hat{\beta}_M +(X_M^TWX_M)^{-1}  \frac{\partial}{\partial \beta_M} \ell_M(\beta_M) \biggl|_{\beta_M=\hat{\beta}_M} \\
&= \hat{\beta}_M + \lambda (X_M^TWX_M)^{-1} s_M
\end{aligned}
$$
where
$$
\ell_M(\beta_M) = \ell \begin{pmatrix} \beta_M \\ 0 \end{pmatrix}
$$
is the log-likelihood of the selected model and
$$
I_M(\hat{\beta}_M) = X_M^TWX_M = X_M^TW(\hat{\beta}_M)X_M 
$$
is its Fisher information matrix evaluated at $\hat{\beta}_M$.
We see that $\bar{\beta}_M$ is defined by one Newton-Raphson step in the selected model from $\hat{\beta}_M$. 

If we had not used the data to select variables $M$ and signs $s_M$, then assuming the
model with variables $M$ is correctly specified, as well as standard
assumptions on $X$ (\cite{bunea2008}),  $\bar{\beta}_M$ would be asymptotically Gaussian centered around
$\beta^*_M$ with approximate covariance $(X_M^TWX_M)^{-1}$. This approximation is of course the usual
one used in forming Wald tests and confidence intervals in generalized linear models.

\section{A more general form and an asymptotic justification}
\label{sec:theory}

We assume that $p$ is fixed, that is, our results for not apply in the high-dimensional regime where $p \rightarrow \infty$.
To state our main result, we begin by considering a general lasso-penalized problem.
Given a  log-likelihood $\ell(\beta)$, denote a $\ell_1$-penalized estimator by
\begin{equation}
\label{eq:likelihood:lasso}
\hat{\beta} = \hat{\beta}_{\lambda} = \text{argmin}_{\beta} [-\ell(\beta)]+ \lambda \|\beta\|_1.
\end{equation}

On the event $\{(\hat{M},s_{\hat{M}})=(M,s_M)\}$, the active block of the  KKT conditions 
are
$$
\frac{\partial}{\partial \beta_M}\ell_M(\beta_M) \biggl|_{\beta_M=\hat{\beta}_M} = \lambda s_M
$$
where $\ell_M$ is the   log-likelihood of the submodel $M$.
The corresponding one-step estimator is
\begin{equation}
\label{eq:onestep}
\begin{aligned}
\bar{\beta}_M &=\hat{\beta}_M  +\lambda  I_M(\hat{\beta}_M)^{-1} s_M \\
 &=\hat{\beta}_M  +  I_M(\hat{\beta}_M)^{-1} \frac{\partial} {\partial \beta_M}  \ell_M(\beta_M) \biggl|_{\beta_M=\hat{\beta}_M}.
\end{aligned}
\end{equation}
where $I_M(\hat{\beta}_M)$ is the $|M| \times |M|$ observed Fisher information matrix of the submodel $M$ evaluated at 
$\hat{\beta}_M$.

In the previous section, we noted that $\bar{\beta}_M$ {\em almost} solves the KKT conditions
for the unpenalized logistic regression model. We further recognized it
as a one step estimator with initial estimator $\hat{\beta}_M$ in the logistic regression model.
In the context \eqref{eq:likelihood:lasso} we have directly defined $\bar{\beta}_M$ as a one-step estimator with the initial
estimator $\hat{\beta}_M$. As long as $\lambda$ is selected so that $\hat{\beta}_M$ is 
$\sqrt{n}$ consistent (usually satisfied by taking $\lambda \propto n^{1/2}$ at least in the fixed $p$ setting considered here)
the estimator $\bar{\beta}_M$ would typically have the same limiting distribution as the unpenalized MLE in the selected model if we had not
used the data to choose the variables to be included in the model.
That is, if we had not selected the variables based on the data,
standard asymptotic arguments yield
\begin{equation}
\bar{\beta}_M \approx N\left(\beta_M^*, I_M(\hat{\beta}_M)^{-1}\right)
\label{eq:bbar}
\end{equation}
where $I_M(\hat{\beta}_M)^{-1}$ is the ``plug-in'' estimate of the asymptotic covariance
of $\bar{\beta}_M$, with the population value being $E_{F}[I_M(\beta_M^*)]^{-1}$. Implicit in this notation is that
$I_M = I_{M,n}$, i.e. the information is based on a sample of size $n$ 
from some model. We specify this model
precisely in Section \ref{sec:model} below.

However, selection with the LASSO has imposed the ``active'' constraints, i.e. we have
observed the following event is true
\begin{equation}
\left\{\text{diag}(s_M)\left[\bar{\beta}_M - I_M(\hat{\beta}_M)^{-1} \lambda s_M)\right] \geq 0 \right\},
\label{eq:onestep-constraints}
\end{equation}
as well as some ``inactive'' constraints that we return to shortly.
Selective inference 
\cite{fithian14:_optim_infer_after_model_selec, TTPNAS2015}
modifies the pre-selection distribution by conditioning on the LASSO having chosen these variables and signs, i.e. by conditioning
on this information we have learned about the data. 

\subsection{Specification of the model}
\label{sec:model}

As the objective function involves a log-likelihood, there is some
parametric family of distributions that is natural to use for inferential
purposes. 
In the
generalized linear model setting, these distributions are models for the laws
$y_i | X$. Combining this with a marginal distribution for $X$ yields a
full specification of the joint law of $(y,X)$. In our only result
below, we consider
the $X_i$'s to be IID draws from some distribution $F$ and the law $y_i|X$ to be independently drawn according to the log-likelihood corresponding to the
generalized linear model setting. In this case, our model
is specified by a pair $(\beta, F)$ and we can now consider asymptotic
behavior of our procedure sending $n \to \infty$. Similarly, for $p$
fixed and any $M \subset \{1, \dots, p\}$ our selected model
is specified by the pair $(\beta_M, F)$ and we can consider similar
asymptotic questions. As is often the case, the most interesting
asymptotic questions  are local alternatives in which $\beta_M$ itself
depends on $n$, typically taking the form $\beta^*_{M,n} = n^{-1/2}\theta_M^*$.
These assumptions are similar to those studied in 
\citet{bunea2008}.  
In local coordinates, \eqref{eq:bbar} could more properly be restated as
\begin{equation}
\bar{\theta}_M = n^{1/2} \bar{\beta}_M \approx N\left(\theta_M^*, n I_M(\hat{\beta}_M)^{-1}\right)
\label{eq:bbar:local}
\end{equation}
where $nI_M(\hat{\beta}_M)^{-1}$ will have non-zero limit 
$$\lim_{n \to \infty}E_{F}[I_M(n^{1/2}\theta_M^*)/n]^{-1}.$$

For example, in the Bernoulli case (binary $Y$),  for any $M \subset \{1, \dots, p\}$ and sample size $n$ 
our selected model is therefore
parametrized by
$(\theta_M^*, F)$ where features are drawn IID according to 
$F$ and, conditionally on $X_i$ we have
$$
y_i | X_i \sim \text{Bernoulli}(\pi(n^{-1/2} \theta_M^*)).
$$

\subsection{Asymptotics of the one-step estimator}

In this section we lay out a description of the limiting conditional distribution of the one-step estimator
in the logistic case. Under our local alternatives, in the selected model the data generating mechanism
is completed determined by the tuple $(n, \theta_M^*, F)$ where $F$ is the distribution of the 
features $X$ and $y|X$ is assumed to follow the parametric logistic regression model with features $X_M$ and 
parameters $\beta_M^*=n^{-1/2}\theta_M^*$. Therefore, any statement about consistency and weak convergence
that follows is a statement about this sequence of data generating mechanisms.

The event \eqref{eq:onestep-constraints} can be rewritten
as 
\begin{equation}
\left\{\text{diag}(s_M)\left[\bar{\theta}_M - (n^{-1/2}\lambda) {\cal I}_M^{-1} s_M)\right]  + R_{M} \geq 0 \right\}
\label{eq:population-constraints}
\end{equation}
where 
$$
\begin{aligned}
R_{M}
&= (n^{-1/2} \lambda) \cdot\left ({\cal I}_M^{-1} - (I_M(\hat{\beta}_M)/n)^{-1}\right) s_M \\
&= (n^{-1/2} \lambda) \cdot\left ({\cal I}_M^{-1} - E_{F}[I_M(n^{-1/2}\theta^*_M)/n]^{-1}\right) s_M  + \\
& \qquad (n^{-1/2} \lambda) \cdot\left (E_{F}[I_M(n^{-1/2}\theta^*_M)/n]^{-1} - (I_M(n^{-1/2}\theta^*_M)/n)^{-1}\right) s_M  + \\
& \qquad (n^{-1/2} \lambda) \cdot D_M(n^{-1/2}\tilde{\theta}^*_M) (n^{-1/2} \theta^*_M - \hat{\beta}_M) s_M
\end{aligned}
$$ is an unobservable remainder. The second equality is just Taylor's theorem with $D_M$ denoting
the derivative
of $I_M/n^{-1}$ with respect to $\beta$ which is evaluated at some $n^{-1/2}\tilde{\theta}^*_M$ between $n^{-1/2}\theta^*_M$ and $\hat{\beta}_M$. If $\lambda = C n^{1/2}$ then all terms
in the above event have non-degenerate limits as $n \to \infty$ with the only
randomness in the event being $\bar{\theta}_M$ and the remainder $R_{M}$.

Pre-selection, the first term of the remainder is seen to converge to 0 by the assumption that the information converges.
The second term is seen to converge to 0 by the strong law of large numbers and the third term is seen to converge to 0 when $\hat{\beta}_M$ is consistent for $n^{-1/2}\theta^*_M=\beta^*_M$.
As we are interested in the selective distribution we need to ensure that this remainder goes to 0 in probability, conditional
on the selection event. For this, it suffices to assume that the probability of selecting variables $M$ is bounded below,
ensuring that Lemma 1 of 
\cite{TT2015}
is applicable to transfer consistency pre-selection to consistency after selection.
In terms of establishing a limiting distribution for inference, we appeal
to the CLT which holds pre-selection and consider its behavior after selection.
It was shown in 
\cite{TT2015}
that
CLTs that hold before selection extend to selective inference after randomization under suitable assumptions.

We provide a sketch of such a proof in our setting.
As we want to transfer a CLT pre-selection to the selective case,
we assume that $\bar{\theta}_M$ satisfies a CLT pre-selection 
under our sequence of data generating mechanisms.
Now, consider the selection event after removing
the ignorable remainder $R_M$ under the assumption that
$\lambda=Cn^{1/2}$
\begin{equation}
\left\{\text{diag}(s_M)\left[\bar{\theta}_M - C \cdot {\cal I}_M^{-1} s_M)\right]\geq 0 \right\}
\label{eq:noremainder}
\end{equation}
If the probability of \eqref{eq:noremainder} converges to some non-negative limit under our sequence of
data generating mechanisms, it must agree with the same probability computed
under the limiting Gaussian distribution. A direct application of the Portmanteau theorem establishes that the sequence of 
conditional distributions of $\bar{\theta}_M$ will therefore converge weakly and this limit will be the 
limiting Gaussian conditioned on \eqref{eq:noremainder}.

This simple argument establishes weak convergence of the conditional distribution for a particular $(\theta_M^*,F)$ sending $n \to \infty$. 
For full inferential
purposes, this pointwise weak convergence is not always sufficient. See \cite{TRTW2015} for some discussion of this topic and {\em honest} confidence intervals. 
A more rigorous treatment of transferring a CLT pre-selection
to the selective model is discussed in 
\cite{TT2015},
where quantitative
bounds are derived to compare the true distribution of a pivotal quantity
to its distribution under the limiting Gaussian distribution.


Let's look at the inactive constraints. In the logistic regression example, with
$\pi=\pi_M(\beta^*_M)$, we see that by construction
$$
\begin{aligned}
X_{-M}^TW(z - X_M\hat{\beta}) &= X_{-M}^T(y - \pi_M(\hat{\beta}_M)) \\
&= X_{-M}^T(y - \pi_M(\beta^*_M)) - E_{F}[X_{-M}^TW(\beta_M^*)X_M](\bar{\beta}_M - \beta^*_M) + \Delta(F,M) s_M + R_{M,2}\\
\end{aligned}
$$
where 
$$
\Delta(F,M) = E_{F}[X_{-M}^TW(\beta_M^*)X_M] E_{F}[ (X_M^TW(\beta^*_M)X_M)]^{-1}
$$
is a population version of the matrix appearing in the well-known
irrepresentable condition [\citet{wainwright2009,tropp2014}]
and the remainder $R_{M,2}$ also going to 0 in probability after appropriate rescaling 
\cite{TT2015}.
Hence, our ``inactive constraints'' can be rewritten in terms of 
the random vector 
$$
X_{-M}^T(y - \pi_M(\beta_M^*)) - E_{F}[X_{-M}^TW(\beta_M^*)X_M](\bar{\beta}_M - \beta^*_M),
$$
the remainder $R_{M,2}$ and a constant vector.
Now, under our selected model, standard asymptotic
arguments show that the random vector 
\begin{equation}
\begin{pmatrix}
\bar{\beta}_M - \beta_M^* \\
X_{-M}^T(y - \pi) - E_{F}[X_{-M}^TW(\beta_M^*)X_M](\bar{\beta}_M - \beta^*_M)
\end{pmatrix}
\label{eqn:standard}
\end{equation}
satisifes a CLT
before selection. It is straightforward to check that under this limiting
Gaussian distribution these two random vectors are independent. Indeed, in the
Gaussian case, they are independent for every $n$. This implies a
simplification similar to \eqref{eq:lasso:poly} 
occurs asymptotically for the problem \eqref{eq:likelihood:lasso}.
While this calculation was somewhat specific to logistic regression, this asymptotic independence of the two
blocks and simplification of the constraints
also holds when the likelihood in \eqref{eq:likelihood:lasso} is an exponential family and with $\beta$
being the natural parameters.

If we knew $\beta_M^*$ and $F$ we could compute all relevant
constants in the constraints and simply apply the
polyhedral lemma to the limiting Gaussian in the CLT mentioned in the
previous paragraph. This would allow for
asymptotically exact selective inference for the selection
event $\{(\hat{M}, s_{\hat{M}}) = (M, s_M)\}$ by construction of a pivotal quantity
\begin{equation}
\label{eq:asym:pivot}
P(\bar{\theta}_M; \lim_{n \to \infty} nE_{F}[I_M(n^{1/2} \theta_M^*)]^{-1}; A, b)
\end{equation}
as in \eqref{eq:tgauss} , where $A$ and $b$ can be derived from the polyhedral constraints 
\eqref{eq:population-constraints}.
Specifically, 
$A=-\text{diag}(s_M), b=- E_{F}[I_M(n^{1/2}\theta^*_M)/n]^{-1} (n^{-1/2}\lambda) s_M) - n^{1/2} R_{M,1}$.

However, the quantities needed to compute
$ \lim_{n \to \infty} nE_{F}[I_M(n^{1/2} \theta_M^*)]^{-1}$
are unknown, though there
are certainly natural plug-in estimators that would be
consistent without selection. This suggests using a plug-in estimate of variance. In 
\cite{TT2015}
 it was shown that, under mild regularity assumptions, consistent estimates of variance
can be plugged into limiting Gaussian approximations
for asymptotically valid selective inference. 
Hence, to construct a practical algorithm, we apply the polyhedral lemma to the limiting distribution of 
$n^{1/2} \bar{\beta}_M$, with $M, s_M$ fixed and $n\nabla^2 \ell_M (\hat{\beta}_M)$ as a plugin estimate for $n E_{F}[I_M(\beta_M^*)]$. 

Thus we have the following result:

\begin{lemma}
\label{lemma:main}
Suppose that the model described in Section \eqref{sec:model} holds for 
all $n$ and some $(\theta_M^*, F)$ such that the corresponding
population covariance ${\cal I}_M(\theta_M^*)$ is non-degenerate. 
Then, the pivot \eqref{eq:asym:pivot} is asymptotically
$U(0,1)$ as $n \to \infty$ conditioned
on having selected variables $M$ with signs $s_M$. Further, plugging in 
$nI_M(\hat{\beta}_M)^{-1}$ both in the limiting variance 
and in the constraints of \eqref{eq:population-constraints} 
of the pivot is also asymptotically $U(0,1)$.
\end{lemma}

As noted in the introduction, a detailed proof of this result will appear elsewhere.

\medskip
{\bf Remark A.} 
In the Gaussian model,
our one-step estimator has the form 

\begin{equation}
\bar\beta_M=\hat\beta_M+(X_M^TX_M)^{-1}X_M^T(y-X\hat\beta_M)=(X_M^TX_M)^{-1}X_M^T y
\end{equation}
with $\bar \beta_M  \sim N((X_M^TX_M)^{-1}X_M^TX\beta,(X_M^TX_M)^{-1}\sigma^2)$ and constraints
\begin{equation}
\left\{\text{diag}(s_M)\left[\bar{\beta}_M - (X_M^TX_M)^{-1})^{-1} \lambda s_M)\right] \geq 0 \right\}.
\label{eq:onestep-constraints2}
\end{equation}
These are just the usual least squares estimates for the active variables.
We note the strong similarity between the one-step estimator  and the ``debiased lasso'' construction of \cite{zhangconf}, \cite{buhlsignif},  \citet{vdgsignif}, and \citet{montahypo2}. In the context of Gaussian regression, the latter approach uses
\begin{equation}
\hat\beta^d\equiv \hat\beta_M+(1/N)\Theta X^T(y-X\hat\beta_M)
\label{eq:debias}
\end{equation}
where $\Theta$ is an estimate of $(X^TX/N)^{-1}$.  This estimator takes a Newton step in the
full model direction. Our one-step estimator has a similar form to (\ref{eq:debias}), but takes a step only in the active variables, leaving the others at 0. Further, the $(X_M^TX_M/N)^{-1}$ is used as the estimate for  $\Theta$.
The debiased lasso (\ref{eq:debias})  uses a full model regularized estimate of $\Theta$ and ignores the constraints in (\ref{eq:onestep-constraints}).
As pointed out by a referee, the debiased lasso is more complex because it does not assume that the lasso has the screening property,
(i.e. the true nonzero set is included in the estimated nonzero set). Another important difference is that their target of inference for the debiased lasso
is a population parameter, i.e.
is determined before observing the data. This is not the case for our procedure.

\medskip
{\bf Remark B.} 
The conclusions of Result 1 can be strengthened to hold uniformly over compact subsets of $\theta_M^*$ parameters. While we do not pursue this here, such results are stated more formally in \cite{TT2015} in the setting where noise is first added to the data
before model fitting and selection. Lemma 1 is a statement about the conditional distribution of the pivot under selected model. In the Gaussian case, similar results hold unconditionally
for the pivot in the saturated model \cite{TT2014,TRTW2015}.

\medskip
{\bf Remark C.} 
For the Cox model, we define the one-step estimator in a similar fashion. In terms of the appropriate distribution 
for inference, we simply replace the likelihood by a partial likelihood.

\subsection{Unpenalized variables}

It is common to include an intercept in logistic regression and other models, which is typically not penalized in the $\ell_1$ penalty. More generally,shows the   
suppose features $U_{n \times k}$ are to have unpenalized coefficients while those  for $X_{n \times p}$ are to be penalized. This changes
the KKT conditions we have been using somewhat, but not in any material way. We now have $\eta = U\alpha + X\beta$
and the KKT conditions now include a set of conditions for the unpenalized variables, say $U$ 
In the logistic regression case, these read as
\begin{equation}
\label{eq:unpenalized:KKT}
U^TW(z - X_M\hat{\beta}_M - U\hat{\alpha}) = 0.
\end{equation}
The corresponding one-step estimator is
$$
\begin{pmatrix}
\bar{\alpha} \\
\bar{\beta}_M
\end{pmatrix}
= \begin{pmatrix}
\hat{\alpha} \\
\hat{\beta}_M
\end{pmatrix}
 + I_M(\hat{\alpha}, \hat{\beta}_M)^{-1} \begin{pmatrix} 0 \\ \lambda s_M \end{pmatrix}
$$
where $I_M$ is the $(|M|+k) \times (|M|+k)$ Fisher information matrix of the submodel $M$.
In terms of constraints, we only really need consider the signs of the selected variables and the corresponding
``plug-in'' form of the active constraints are
$$
\text{sign}\left(\bar{\beta}_M - E_M\left(I_M(\hat{\alpha}, \hat{\beta}_M)^{-1} \begin{pmatrix} 0 \\ \lambda s_M \end{pmatrix}\right)\right) = s_M.
$$
The population version uses the expected Fisher information at $(\alpha^*,\beta_M^*)$ instead of the observed information and 
$E_M$ is the matrix that selects rows corresponding to $M$.
As our one-step estimator is expressed in terms of the likelihood this estimator can be used in problems
that are not regression problems but that have unpenalized parameters such as the graphical LASSO discussed in Section \ref{sec:glasso}.

\subsection{The random $X$-case}
\label{sec:randomX}
The truncated Gaussian theory  of  
\citet{lee2014} assumes that $X$ is fixed, and conditions on it in the inference.  When $X$ is random (most often the case),
this ignores its inherent variability  and makes the inference non-robust when the error variance in non-heterogeneous.
This point is made forcefully by \citet{buja2016}.  

The one-step estimation framework of this paper provides a way deal with the problem. Consider for simplicity the Gaussian case for the lasso of \cite{lee2014}, which expresses the selection
as $Ay \leq 0$  with $y \sim N(\mu, I \sigma)$.  Above, we have re-expressed this as $\bar \beta_M \sim N(\beta^*,  \Sigma)$ where $\bar \beta_M $ is the one-step estimator
for the selected  model.
In the Gaussian case, $\bar \beta_M$ is just $\hat \beta_M$, the usual least squares estimate on the selected set and $\Sigma=(X_M^TX_M)^{-1} \sigma^2$.
Now analogous to (\ref{eqn:standard}), for the Gaussian case we have the asymptotic result

\begin{equation}
\begin{pmatrix}
\bar{\beta}_M - \beta_M^* \\
X_{-M}^T(y - X \beta^*) - E_{F}[X_{-M}^TX_M](\bar{\beta}_M - \beta^*_M) \
\end{pmatrix}
\sim N(0,\Sigma')
\label{eqn:gaussfull}
\end{equation}
This suggests that we can use the pairs bootstrap to estimate the {\em unconditional} variance-covariance matrix $\Sigma'$ and then simply apply the
polyhedral lemma, as before. Alternatively, a sandwich-style estimator of $\Sigma'$ can be used.

\begin{figure}
\includegraphics[width=\textwidth]{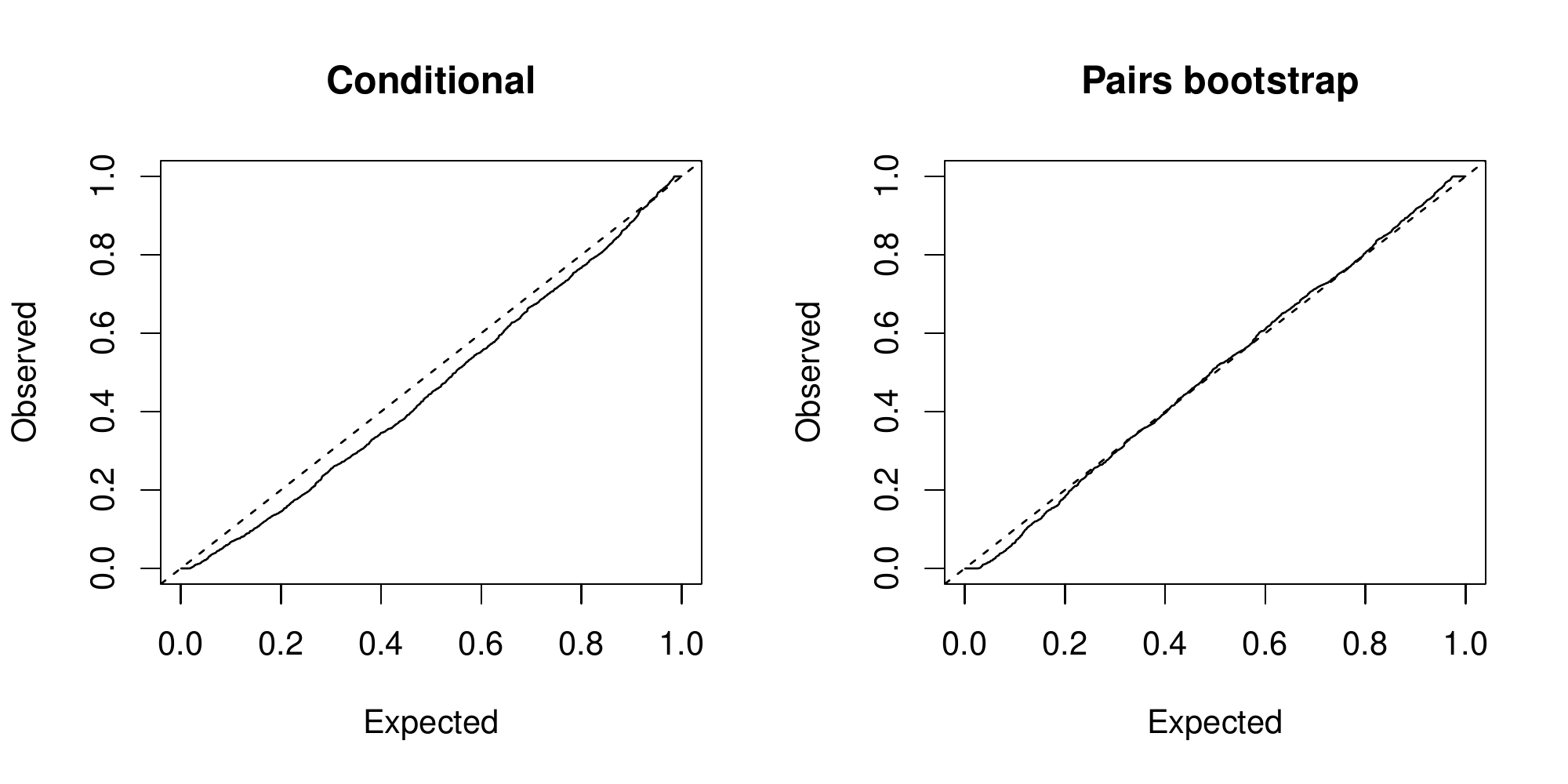}
\caption{\em P-values for the lasso in the Gaussian setting:  $n=200, p=50$, 20 strong signals. Predictors have pairwise correlation of 0.3 and variance of the errors depends on non-signal variables. Shown are the quantile-quantile  plots
for the non-signal variables in realizations for which the lasso has successfully screened (captured all of the signal variables).
We see that  the conditional analysis yields anti-conservative p-values while the pairs bootstrap gives p-values closer to uniform.}
\label{fig:pairs}
\end{figure}

Figure \ref{fig:pairs} shows an example, illustrating how the pairs bootstrap can give robustness again heterogeneity of the error variance.
Details are in the caption.

\section{Simulations}
\label{sec:simulations}
\begin{figure}[htp]
\begin{center}
\includegraphics[width=\textwidth]{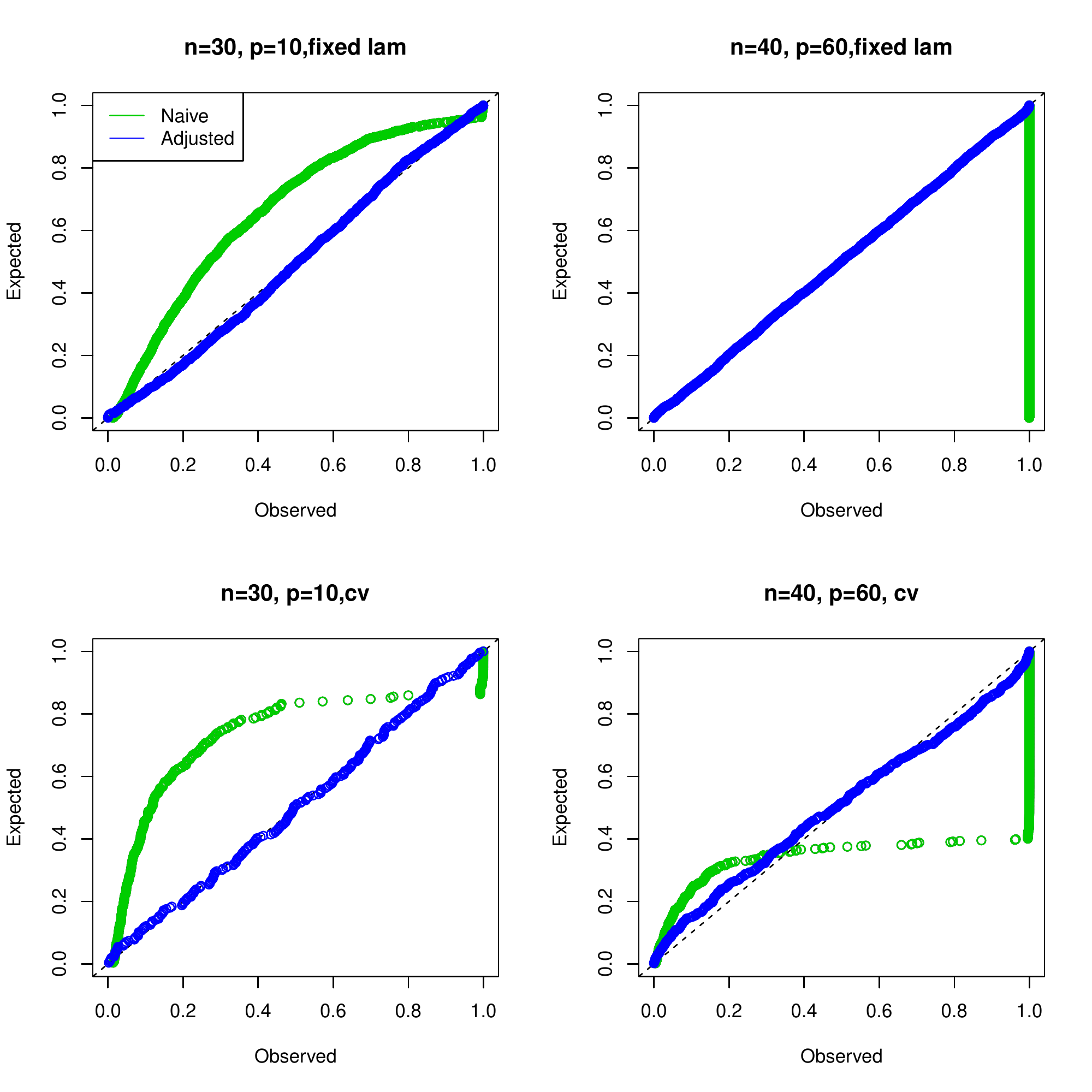}
\end{center}
\caption{\em P-values for the logistic regression model, null setting. The top panels use a fixed $\lambda$ while the bottom ones use cross-validation to choose $\lambda$.
}
\label{fig:sim1}
\end{figure}

\begin{figure}[htp]
\begin{center}
\includegraphics[width=\textwidth]{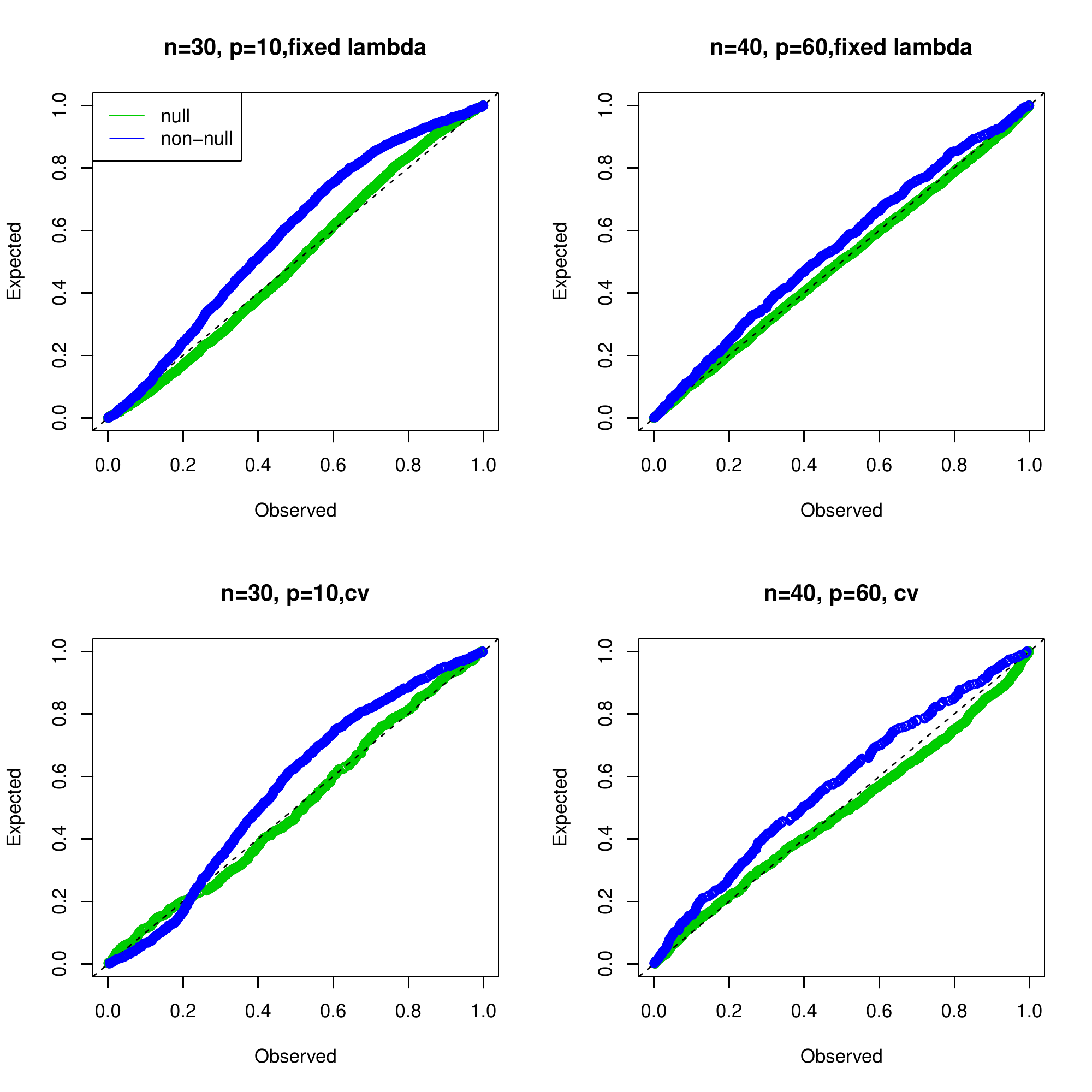}
\end{center}
\caption{\em P-values for the  logistic regression model, non-null setting. The top panels use a fixed $\lambda$ while the bottom ones use cross-validation to choose $\lambda$.
}
\label{fig:sim2}
\end{figure}

To assess performance in the $\ell_1$-penalized logistic model, we generated Gaussian features with pairwise correlation 0.2 in two scenarios: $n=30, p=10$ and $n=40, p=60$.
Then $y$ was  generated as ${\rm Pr}(Y=1|x)=1/(1+\exp(-x^T\beta))$. There  are two signal settings: null ($\beta=0$)  (Figure \ref{fig:sim1})
non-null   ($\beta=(5,0,0\ldots )$  (Figure \ref{fig:sim2}).  Finally, in each case we tried two methods for choosing the regularization parameter $\lambda$:
a  fixed value yielding a moderately sparse model and cross-validation. The Figures show the cumulative distribution function of the resulting
p-values over 1000 simulations. Thus a function above the 45 degree line indicates an anti-conservative test in the null setting and a test with some
power in the non-null case.
We see the adjusted p-values are close to uniform under the null  in every case and show power in the non-null setting.
Even with cross-validation   -based choice for $\lambda$ the type I error seems to be controlled, although  we have no theoretical support for this finding.
In Figure \ref{fig:sim1} we also plot the naive p-values from  GLM theory: as expected they are very anti-conservative. 

\begin{figure}[htp]
\begin{center}
\includegraphics[width=\textwidth]{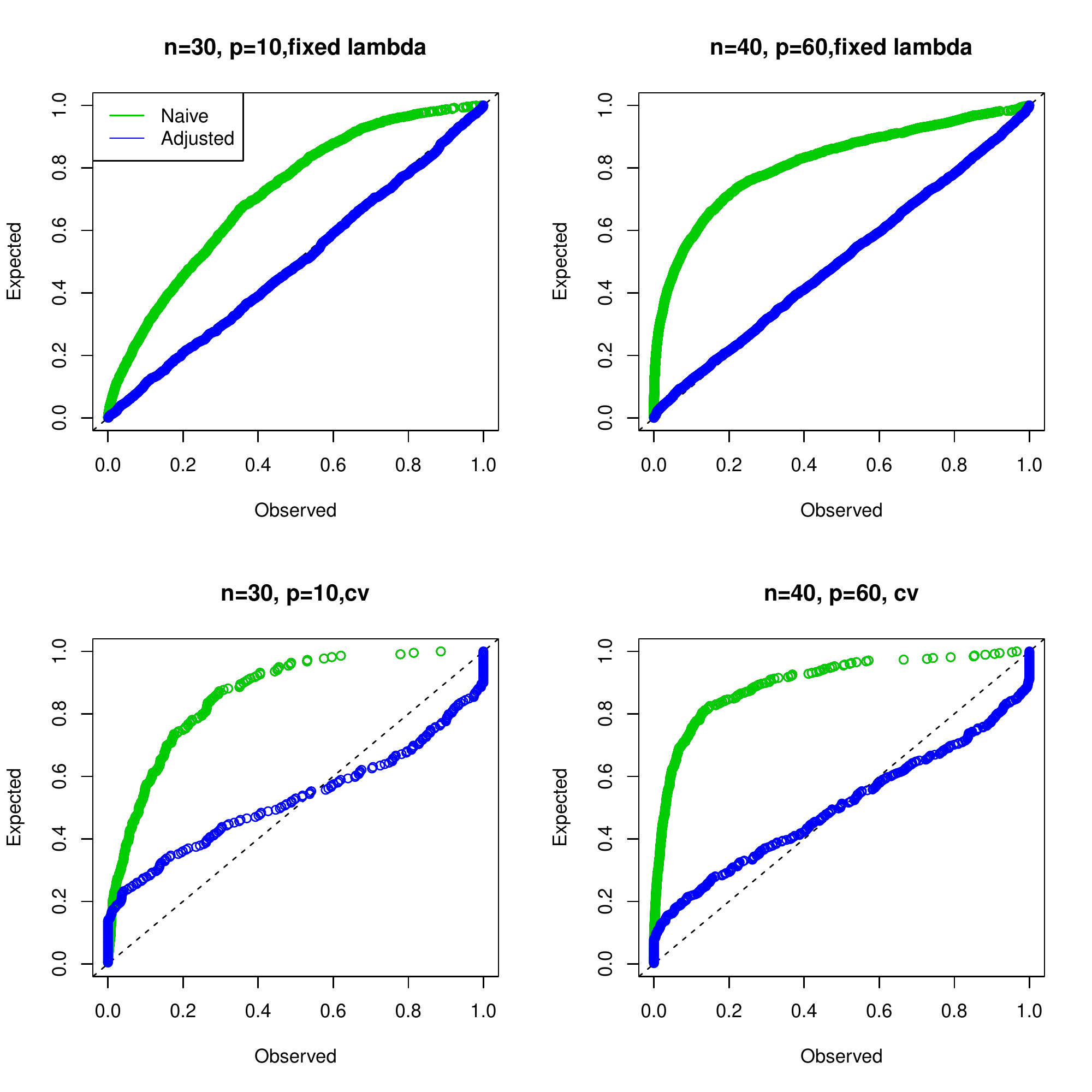}
\end{center}
\caption{\em P-values for the Cox Model, null setting. The top panels use a fixed $\lambda$ while the bottom ones use cross-validation to choose $\lambda$.
}
\label{fig:sim3}
\end{figure}

Figure \ref{fig:sim3} shows the results of an analogous experiment for the Cox model in the null setting, using  exponential survival times
and random 50\% censoring.  Type I error control is good, except  in the cross-validation case where it is badly anti-conservative for
smaller p-values. We have seen similar behavior in the Gaussian lasso setting, and this phenomenon deserves further study.

Table \ref{tab:boot1}
shows the miscoverage and median lengths of intervals for the logistic regression example in the null setting, with a target miscoverage of 10\%.
The intervals can sometimes be very long, and in fact, have infinite expected length.

\begin{table}[!h]
\centering
\begin{small}
\begin{tabular}{ccccc}
  \hline
 & N$>$p, fixed $\lambda$ & N$>$p, cv & N$<$p, fixed $\lambda $& N$<$p, cv \\ 
  \hline
 miscoverage & 0.11 & 0.11 & 0.14 & 0.14 \\ 
 median length & 8.07 & 6.63 & 5.75 & 29.69 \\
   \hline
\end{tabular}
\end{small}
    \caption[tab:boot1]{\em Lasso-penalized logistic regression. Details as in Figure \ref{fig:sim2}. Shown are the  the miscoverage  and median length of  selection (confidence)  intervals
for main proposal of this paper.  The target miscoverage is 10\%.
Selection of $\lambda$ is done using either a fixed value yielding moderate sparsity or cross-validation (cv).}
\label{tab:boot1}
\end{table}

As a comparison, Table \ref{tab:boot2}  shows
analogous results for Gaussian regression, using
the proposal of \citet{lee2014}.
For estimation of the error variance $\sigma^2$, we used the mean residual error for $N>p$ and the cross-validation estimate of
\cite{RTF2013} for $N<p$.  
\begin{table}[!h]
\centering
\begin{small}
\begin{tabular}{ccccc}
  \hline
 & $N>p$ , fixed $\lambda $ &$ N>$p, cv & $N<p$, fixed $\lambda$ & $N<p$, cv \\ 
  \hline
miscoverage & 0.11 & 0.12 & 0.13 & 0.10 \\ 
median length& 12.29 & 6.90 & 20.50 & 32.06 \\ 
 \hline
\end{tabular}
\end{small}
\caption[tab:boot2]{\em Lasso-penalized Gaussian regression.  Shown are  the  miscoverage  and median lengths of  selection (confidence) intervals
based on the Gaussian model of  \citet{lee2014}.  The target level is 10\%.  Selection of $\lambda$ is done using either a fixed value yielding moderate sparsity or cross-validation (cv).}
\label{tab:boot2}
\end{table}

Again, the intervals can be quite long. There are potentially better ways to construct the intervals:
\cite{TRTW2015} propose  a bootstrap method for post-selection inference that in our current problem would
draw bootstrap samples from $z_1, z_2, \ldots z_n$ and use  their empirical distribution  in the polyhedral lemma (in place
of the Gaussian distribution). The ``randomized response''  strategy provides another way to obtain shorter intervals, at the expense of increased computation: see \citet{TT2015}.

\section{Examples}
\label{sec:examples}

\subsection{Liver data}
The data in this example and the following  (edited) description were provided by D. Harrington and
T. Fleming.

``Primary biliary cirrhosis (PBC) of the liver is a rare but fatal chronic liver disease of unknown cause,
with a prevalence of about 50-cases-per-million population. The
primary pathologic event appears to be the destruction of interlobular
bile ducts, which may be mediated by immunologic mechanisms.

The following briefly describes data collected for the Mayo Clinic
trial in PBC of the liver conducted between
January, 1974 and May, 1984 comparing the drug D-penicillamine (DPCA) with a
placebo.
The first 312 cases participated
in the randomized trial of D-penicillamine versus placebo, and
contain largely complete data.  An additional 112 cases
did not
participate in the clinical trial, but consented to have basic
measurements recorded and to be followed for survival.  Six of those
cases were lost to follow-up shortly after diagnosis, so
there are data here on an additional 106 cases as well as
the 312 randomized participants.''

\begin{figure}[ht]
\includegraphics[width=\textwidth]{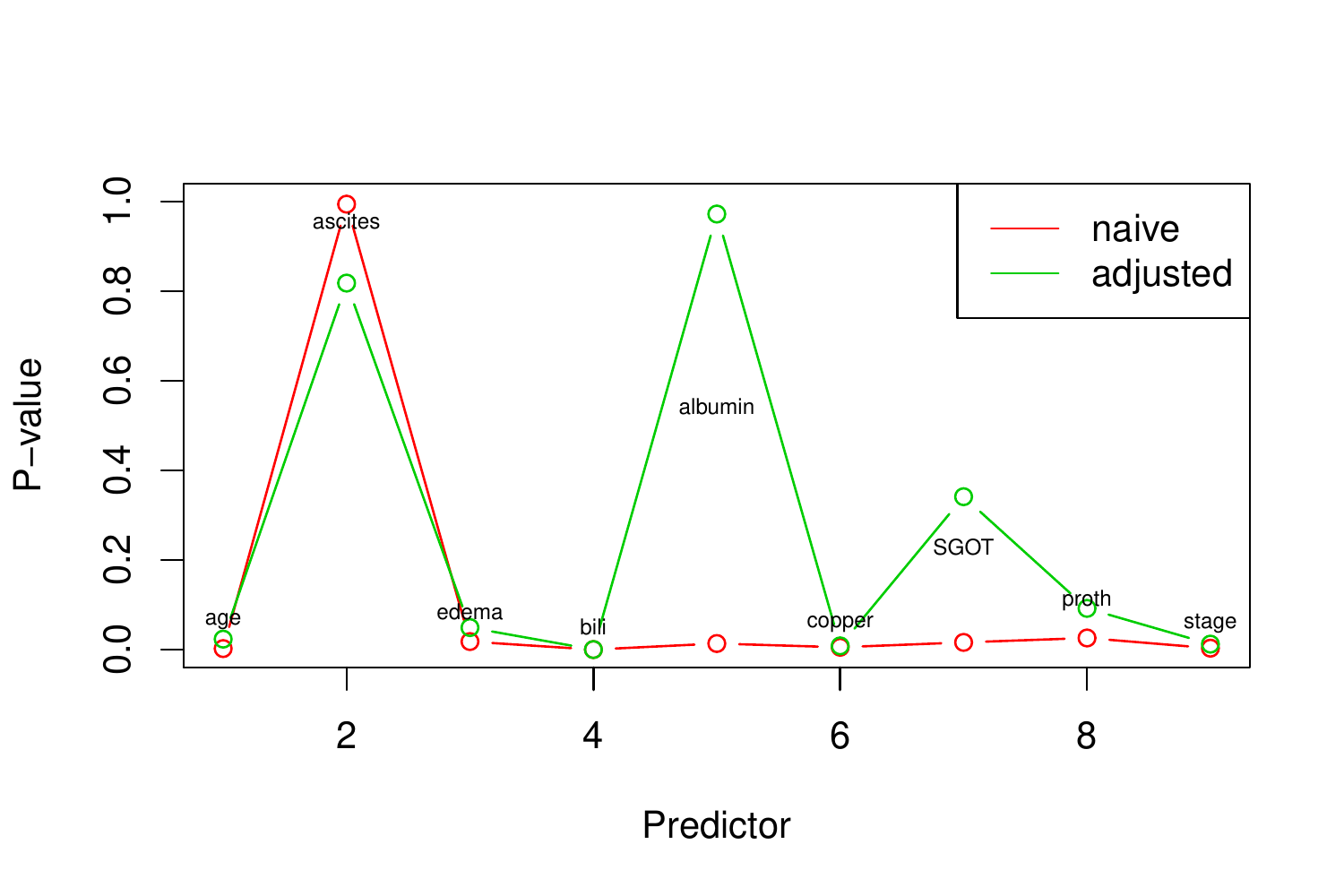}
\caption{\em P-values for  Cox model applied  to the liver data}
\label{fig:liver}
\end{figure}

\begin{figure}[ht]
\includegraphics[width=\textwidth]{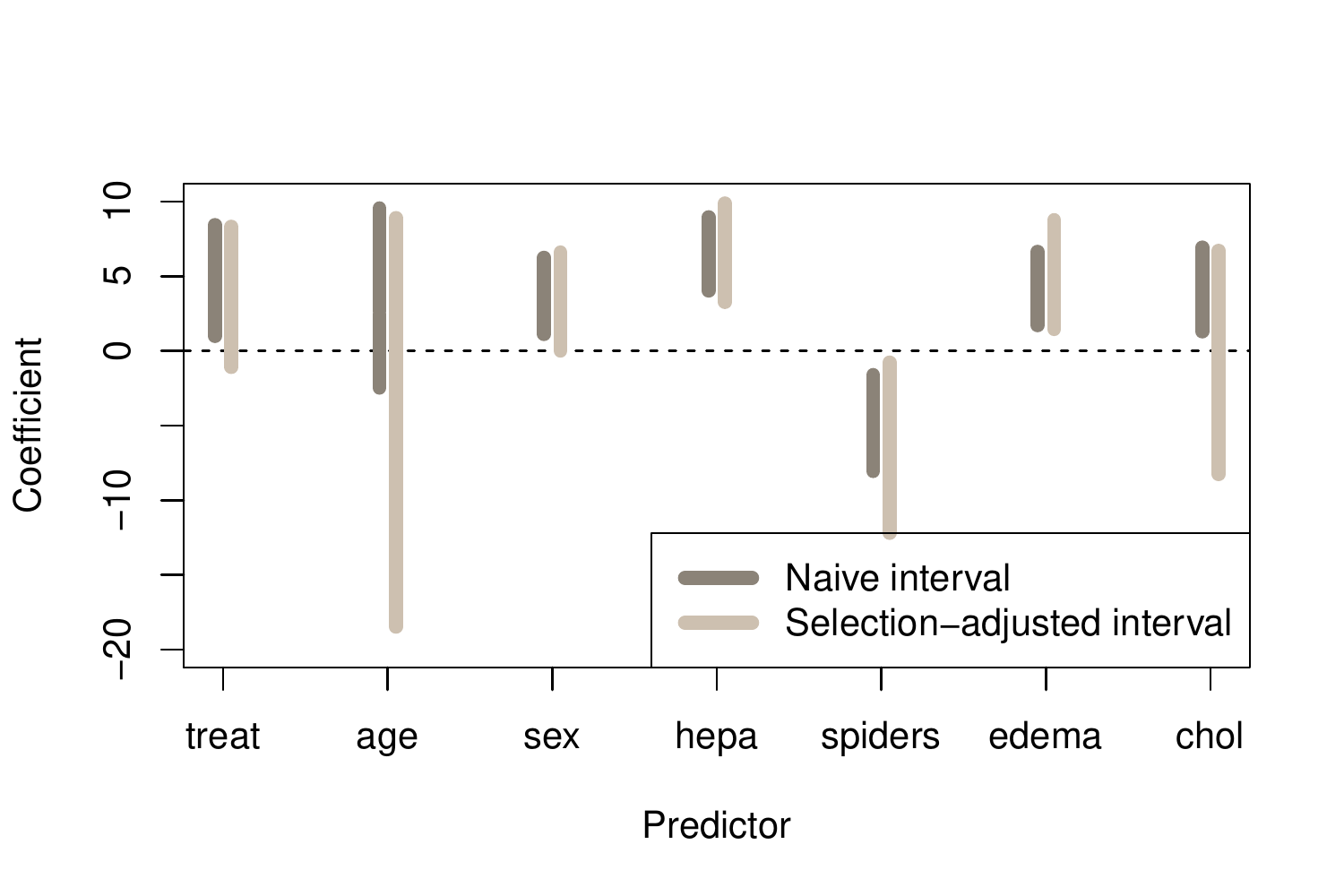}
\caption{\em Selection intervals for Cox model applied to the liver data}
\label{fig:llverSel}
\end{figure}

We discarded observations
with missing values, leaving 276 observations.
The predictors are
\small
\begin{tabbing}
\= $Z_{17}:$  \=Histologic stage of disease, graded 1,2,3, or 4. \kill
\> $X_1:$     \>\parbox[t]{5.1in}{Treatment Code,
                        1 = D-penicillamine, 2 = placebo.}\\
\> $X_2:$     \>\parbox[t]{5.1in}{Age in years. For the first 312 cases, age
                        was calculated by dividing the number of days
                        between birth and study registration by 365.}\\
\> $X_3:$     \>Sex, 0 = male, 1 = female.  \\
\> $X_4:$     \>Presence of ascites, 0 = no, 1 = yes.  \\
\> $X_5:$     \>Presence of hepatomegaly, 0 = no, 1 = yes.  \\
\> $X_6:$     \>Presence of spiders, 0 = no, 1 = yes.  \\
\> $X_7:$     \>\parbox[t]{5.1in}{Presence of edema, 0 = no, .5 yes but responded
                        to diuretic treatment, 1 = yes, did
                        not respond to treatment.}  \\
\> $X_8:$     \>Serum bilirubin, in mg/dl.  \\
\> $X_9:$     \>Serum cholesterol, in mg/dl.  \\
\> $X_{10}:$  \>Albumin, in gm/dl.  \\
\> $X_{11}:$  \>Urine copper, in $\mu$g/day.  \\
\> $X_{12}:$  \>Alkaline phosphatase, in U/liter.  \\
\> $X_{13}:$  \>SGOT, in U/ml.  \\
\> $X_{14}:$  \>Triglycerides, in mg/dl.  \\
\> $X_{15}:$  \>\parbox[t]{5.1in}{Platelet count; coded value is number of
                        platelets per cubic ml. of blood divided by
                        1000.}  \\
\> $X_{16}:$  \>Prothrombine time, in seconds.  \\
\> $X_{17}:$  \>\parbox[t]{5.1in}{Histologic stage of disease,
                        graded 1, 2, 3, or 4.}
\end{tabbing}
\normalsize
We applied Cox's proportional hazards model.
Figures \ref{fig:liver} and \ref{fig:llverSel}  show the results. As expected, the adjusted p-values are larger than the naive ones and the
corresponding selection (confidence) intervals tend to be wider.

\section{The Graphical lasso}
\label{sec:glasso}
Another, different, example is the graphical lasso for  estimation of sparse inverse covariance graphs. 
Here  we have   data $X_{n\times p} \sim N(0,\Sigma)$. Let $S=X^TX/N,  \Theta=\Sigma^{-1}$.

We maximize
\begin{equation}
\ell(\Theta)=  \log{\rm det} \Theta- {\rm tr} S\Sigma -\lambda  ||\Theta||_1.
\end{equation}
where the norm in the second is the sum of the absolute values.

The KKT conditions have the form
\begin{eqnarray}
\Theta^{-1}-S-\lambda s(\Theta)=0
\end{eqnarray}
or for one row/column, $\Sigma_{11} \beta-s_{12}-\lambda s(\beta)=0$,
where $\beta$ is a $p-1$ vector in the $p$th row/col of $\Theta$, excluding the diagonal, and  $\Sigma_{11}$ is the block of $\Sigma$ excluding one row and column.
Defining $R=d^2\ell/d\Theta d\Theta^T$ we have
\begin{equation}
\bar\Theta_M=\hat\Theta_M+\lambda R^{-1}s_M
\end{equation}
Hence we apply the polyhedral lemma to  $\bar\Theta_M \sim N(\Theta_M^*,R^{-1})$ with constraints $-{\rm diag}(s_M)(\bar \Theta_M-R^{-1}s_M )\leq  0.$  From this we can obtain p-values for testing whether a link parameter is zero ($H_0: \theta_{jk}=0$  ) and confidence intervals for $\theta_{jk}$.
We note the related work on high-dimensional inverse covariance estimation in \cite{JG2014}.

\begin{figure}[ht]
\begin{center}
\includegraphics[width=.5\textwidth]{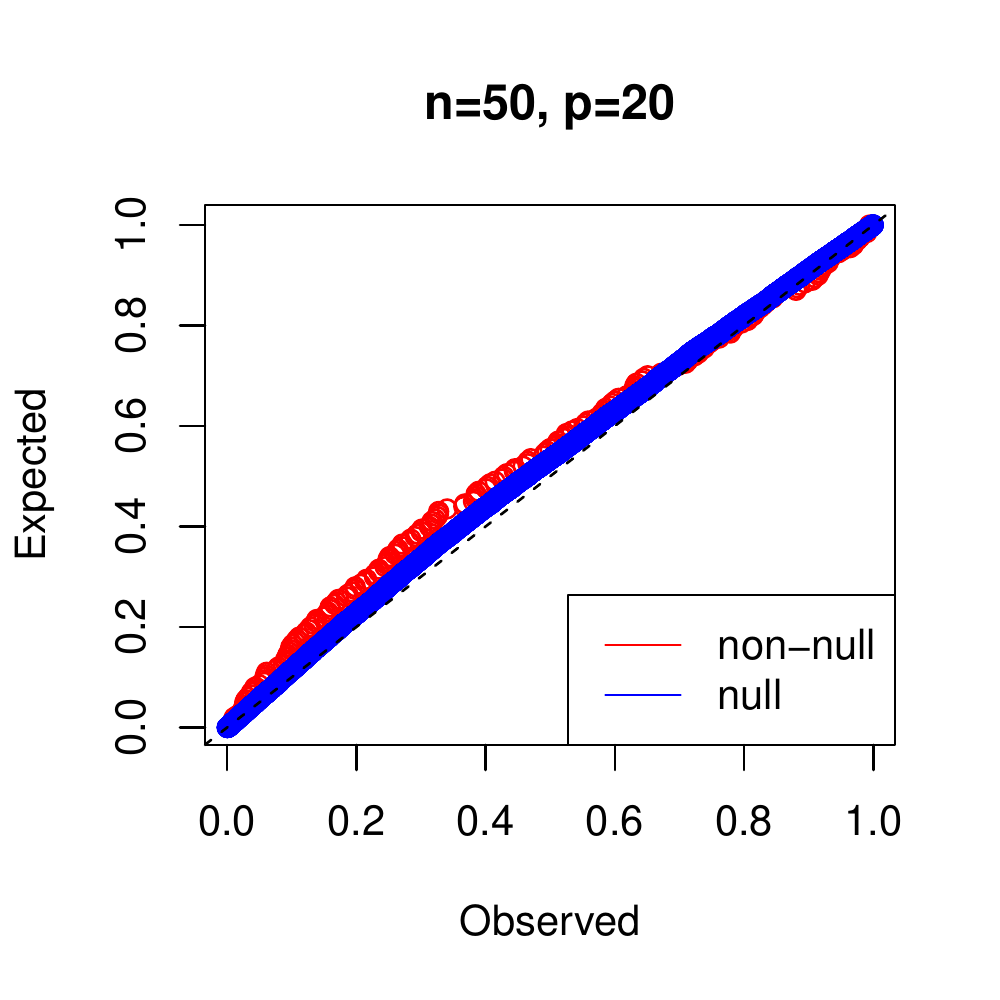}
\end{center}
\caption{\em Simulation results: P-values for the graphical lasso. Details are given in the text.}
\label{fig:sim4}
\end{figure}

Figure \ref{fig:sim4} shows the results of a simulation study with $n=80, p=20$ with the components $X_1, X_2 ,\ldots X_p$ being standard
Gaussian variables. All components were generated independently except for the first two, which had correlation  0.7.
A fixed moderate value of the regularization parameter was used.
Conditioning on realizations for which the  partial correlation for the first two variables was non-zero, the Figure shows the p-values for 
non-null (1,2) entry and the null (the rest). We see that the null p-values are close to uniform and the non-null ones are (slightly) sub-uniform.
       
\subsection{Example}

\begin{figure}[!ht]
\centering
\includegraphics[width=.6\textwidth]{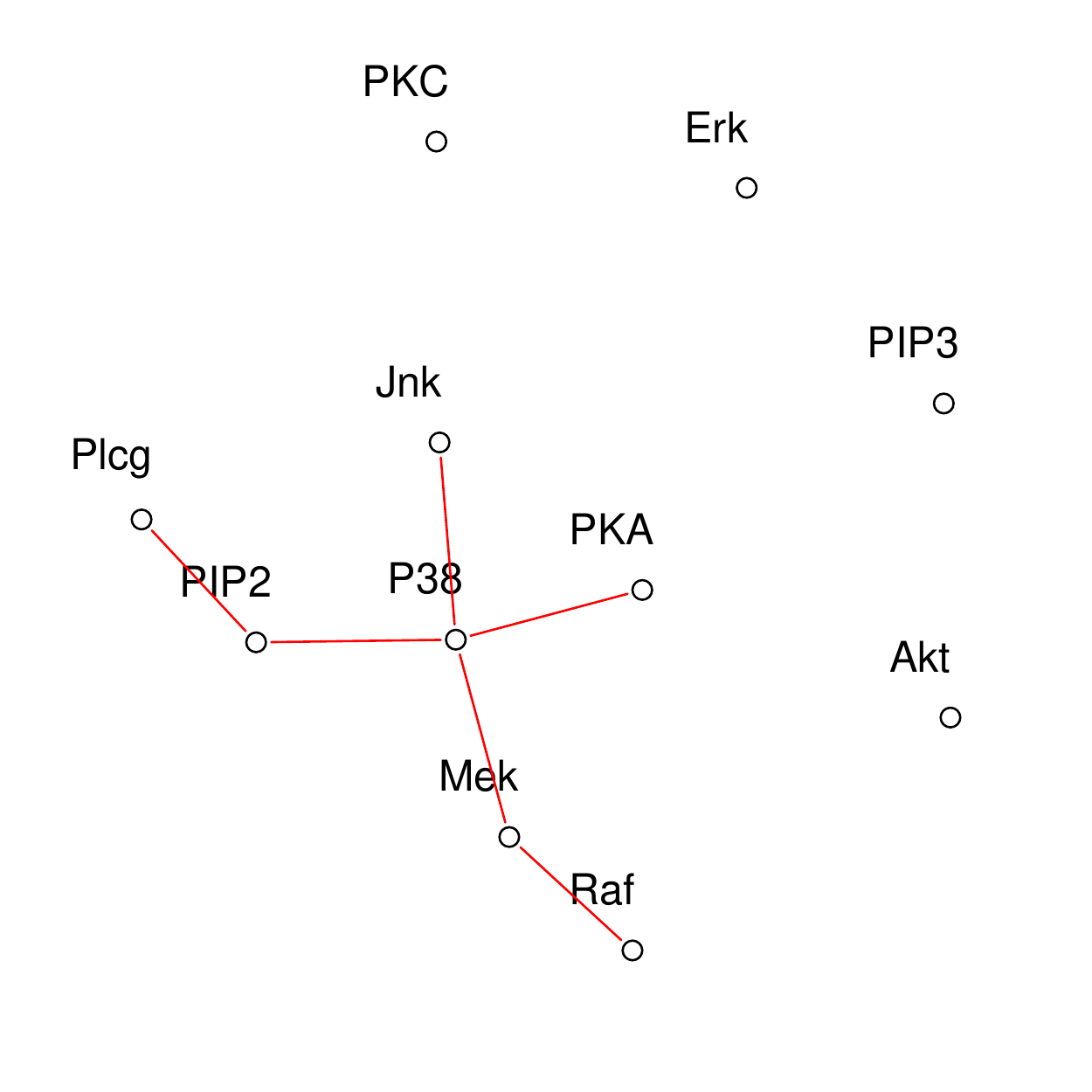}
\caption{\em Results for protein data. The red lines indicate non-zero fitted entries in $\hat\Theta$.}
\label{fig:sachs}
\end{figure}

Here we analyze the protein data discussed in \citet{FriHasTib08}.
The measurements are from flow cytometry, with 11 proteins measured over 7466 cells.
Table \ref{tab:prot} and Figure \ref{fig:sachs} show the results of applying the post-selection procedure with a moderate value of the regularization
parameter $\lambda$. Six interactions are present in the selected model, with only
one (Mek-P38) being strongly significant.

\begin{table}[!h]
\begin{center}
\begin{tabular}{|lr|}
\hline
Protein pair & P-values\\
\hline
Raf -Mek &0.789\\
Mek -P38 &0.005\\
Plcg- PIP2& 0.107\\
PIP2 -P38 &0.070\\
PKA -P38 &0.951\\
P38 -Jnk &0.557\\
\hline
\end{tabular}
\end{center}
\caption{\em P-values from the graphical lasso applied to the protein data}
\label{tab:prot}
\end{table}

\section{Discussion}
\label{sec:discussion}
We have proposed a method for post-selection inference, with applications to $\ell_1$-penalized likelihood models.
These include generalized linear models, Cox's proportional hazards model, and the graphical lasso.
As noted earlier, while our focus has been on selection via $\ell_1$-penalization, a similar approach can be applied to forward stepwise
methods for likelihood models, and in principle, least angle regression.

A major challenge remains in the estimation of the tuning parameter $\lambda$.
One possibility is to use a choice proposed by \citet{negahban2012} (Theorem 1),
which is $\lambda'=2\cdot {\rm E} [{\rm max}_j \nabla L(0)]$.
In the logistic and Cox models, for example, this is not a function of $y$ and
hence the proposals of this paper can be applied. More generally, it would be desirable to allow for the choice of $\lambda$ by cross-validation in our methodology. 
Choosing $\lambda$ by cross-validation is feasible, particularly in the ``randomized response'' setting of \cite{TT2015} though this approach requires
MCMC for inference. A related approach to selective inference after cross-validation is described in \cite{loftus_CV}.

 The proposals of this paper are implemented in  our {\tt selectiveInference} R package in the public  CRAN repository.

\bigskip

{\bf Acknowlegments:} 
\section*{Appendix}
\subsection*{The Polyhedral lemma and truncation limits for post-selection Gaussian inference}

\medskip
Let $y\sim N(\mu,\Sigma)$ and suppose that we apply the lasso with parameter
$ \lambda$ to data $(X, y)$, yielding active variables $M$.
\citet{lee2014} show that for the active variables,
$A_1=-D(X_M^TX_M)^{-1}X_M^T, b_1=-D(X_M^TX_M)^{-1}\lambda s$
where $D={\rm diag(}s)$.
For inactive variables,
$A_0=\frac{1}{\lambda}
\begin{pmatrix}
 X_{-M}^T ) \\
   -X_{-M}^T
   \end{pmatrix}
   $, $b_0= \begin{pmatrix}
              \be+X_{-M}^TX_M\hat\beta/\lambda \\
                 \be-X_{-M}^TX_M\hat\beta/\lambda
              \end{pmatrix}
              $

              Finally, we define $A=\begin{pmatrix} A_1\\ A_0  \end{pmatrix}
, b=(b_1,b_0)$.
They also show that
\begin{equation}
\{A y\leq b\}=\{\Nu^-(y)\leq \gamma^Ty\leq\Nu^+(y),\;\Nu^0(y)\geq 0\},
\label{eq:tn3}
\end{equation}
and furthermore, $\gamma^Ty$ and $(\Nu^-(y),\Nu^+(y),\Nu^0(y))$ are statistically independent.
This surprising result is known as the {\em polyhedral lemma}.\index{polyhedral lemma}
Let $ c\equiv \Sigma\gamma(\gamma^T\Sigma \gamma)^{-1}, r\equiv (I_N-c\gamma^T)y$.
Then the  three values on the righthand side of (\ref{eq:tn3}) are computed via

\begin{equation}
\label{eq:amazing}
  \begin{aligned} 
\Nu^-(r)&=\max_{j:(Ac)_j <0} \frac{b_j-(A r)_j}{(Ac)_j}\\
\Nu^+(r)&=\min_{j:(Ac)_j >0} \frac{b_j-(A r)_j}{(Ac)_j}\\
\Nu^0(r)&=\min_{j:(Ac)_j =0} b_j-(A r)_j.
\end{aligned}
\end{equation}

Hence the selection event $\{A y \leq b\}$ is equivalent to the event that $\gamma^T y$ falls into a certain range, a range depending on $A$ and  $b$.
This equivalence and the independence means that the conditional inference on $\gamma^T\mu$ can be made using the truncated distribution of $\gamma^T y$, a truncated normal distribution.
\subsection*{The Hessian for the graphical lasso}

Let $(\Delta_{ij})_{1 \leq i \leq j \leq p}$ denote the upper triangular parameters for the graphical lasso  and
$$
\Theta(\Delta)_{ij} = 
\begin{cases}
\Delta_{ji} & i > j \\
\Delta_{ij} & i \leq j
\end{cases}
$$
be the symmetric version so that
$$
\frac{\partial}{\partial \Delta_{ij}} \Theta_{kl} = \delta_{ik}\delta_{jl} + \delta_{il} \delta_{jk}.
$$

Now,
$$
\frac{\partial^2}{\partial \Theta_{ij} \partial \Theta_{kl}}(-\log \det \Theta) = \text{Tr}(e_ie_j^T\Theta^{-1} e_ke_l^T \Theta^{-1}) = \Sigma_{jk} \Sigma_{il}
$$
with $\Sigma=\Theta^{-1}$. Note that we evaluate this at a symmetric matrix, i.e. $\Theta^T = \Theta.$

Therefore,
$$
\begin{aligned}
\frac{\partial^2}{\partial \Delta_{ij} \partial \Delta_{kl}}(- \log \det(\Theta(\Delta))) &= \sum_{i',j',k',l'} \Sigma_{j'k'} \Sigma_{i'l'} (\delta_{ii'} \delta_{jj'} + \delta_{ij'} \delta_{ji'}) (\delta_{kk'} \delta_{ll'} + \delta_{kl'} \delta_{lk'}) \\
&= 2( \Sigma_{il} \Sigma_{jk} + \Sigma_{ik} \Sigma_{jl}).
\end{aligned}
$$

\bibliographystyle{agsm}
\bibliography{L1}

\end{document}